\documentclass{emulateapj}
\usepackage{epsfig}
\bibpunct{(}{)}{;}{a}{}{,}

\usepackage{natbib,amsmath}

\begin{document}

\lefthead{PDLA Ly$\alpha$ BLOB}
\righthead{HENNAWI et al.}

\def\qso{SDSS~J1240$+$1455}
\def\name{PDLA/SDSSJ1240$+$1455}
\def\intl{\int\limits}
\def\nstat{$\approx 630$}
\def\perd{\;\;\; .}
\def\cmma{\;\;\; ,}
\def \hkpc      {h^{-1}\,{\rm kpc}}
\def\ltk{\left [ \,}
\def\ltp{\left ( \,}
\def\ltb{\left \{ \,}
\def\rtk{\, \right  ] }
\def\rtp{\, \right  ) }
\def\rtb{\, \right \} }
\newcommand{\snrlim}{SNR$_{lim}$}
\newcommand{\nhi}{$N_{\rm HI}$}
\newcommand{\mnhi}{N_{\rm HI}}
\newcommand{\flls}{f_{\rm HI}^{\rm LLS}}
\newcommand{\fdla}{f_{\rm HI}^{\rm DLA}}
\newcommand{\llls}{$\ell_{\rm LLS}$}
\newcommand{\ldla}{\ell_{\rm DLA}}
\newcommand{\fnhi}{$f_{\rm HI}(N,X)$}
\newcommand{\mfnhi}{f_{\rm HI}(N,X)}
\newcommand{\Nth}{2 \sci{20} \cm{-2}}
\newcommand{\taux}{$d\tau/dX$}
\newcommand{\gz}{$g(z)$}
\newcommand{\omg}{$\Omega_g$}
\newcommand{\ostr}{$\Omega_*$}
\newcommand{\momg}{\Omega_g}
\newcommand{\olls}{$\Omega_g^{\rm LLS}$}
\newcommand{\odla}{$\Omega_g^{\rm DLA}$}
\newcommand{\oneut}{$\Omega_g^{\rm Neut}$}
\newcommand{\olwz}{$\Omega_g^{\rm 21cm}$}
\newcommand{\ndla}{71}
\newcommand{\kms}{km~s$^{-1}$ }
\newcommand{\cm}[1]{\, {\rm cm^{#1}}}
\newcommand{\mkms}{{\rm \; km\;s^{-1}}}
\newcommand{\delv}{\Delta v}
\newcommand{\ohi}{$\Omega_g$}
\newcommand{\lya}{Ly$\alpha$}
\newcommand{\lyb}{Ly$\beta$}
\newcommand{\nv}{N\,V}
\newcommand{\ovi}{O\,VI}
\newcommand{\N}[1]{{N({\rm #1})}}
\newcommand{\sci}[1]{{\rm \; \times \; 10^{#1}}}
\newcommand{\msol}{M_\odot}
\def \cgsflux   {{\rm\,erg\,s^{-1}\,cm^{-2}}}
\def \cgssb {{\rm\,erg\,s^{-1}\,cm^{-2}\,arcsec^{-2}}}

\def\bea{\begin{eqnarray}}
\def\eea{\end{eqnarray}}
\def\be{\begin{equation}}
\def\ee{\end{equation}}

\title{A $z=3$ \lya\ Blob Associated with a Damped \lya\ System
  Proximate to its Background Quasar}

\author{
  Joseph F. Hennawi\altaffilmark{1,2},
  J. Xavier Prochaska\altaffilmark{3}, 
  Juna Kollmeier\altaffilmark{4,5}, 
  Zheng Zheng\altaffilmark{6,7}}

\altaffiltext{1}{Department of Astronomy, 601 Campbell Hall, 
  University of California, Berkeley, CA 94720-3411}
\altaffiltext{2}{NSF Astronomy and Astrophysics Postdoctoral Fellow}
\altaffiltext{3}{Department of Astronomy and Astrophysics, 
  UCO/Lick Observatory;
  University of California, 1156 High Street, Santa Cruz, CA  95064;
  xavier@ucolick.org}
\altaffiltext{4}{Observatories of the Carnegie Institution of Washington, 
  813 Santa Barbara Street, Pasadena, CA 91101.}
\altaffiltext{5}{Hubble Fellow,Carnegie-Princeton Fellow}
\altaffiltext{6}{Institute for Advanced Study, School of Natural Sciences, 
Princeton, NJ 08540}
\altaffiltext{7}{John Bahcall Fellow}

\begin{abstract}
  We report on the discovery of a bright \lya\ blob associated with
  the $z=3$ quasar SDSS~J124020.91$+$145535.6 which is also coincident
  with strong damped \lya\ absorption from a foreground galaxy (a
  so-called proximate damped \lya\ system; PDLA).  The one dimensional
  spectrum acquired by the Sloan Digital Sky Survey (SDSS) shows a
  broad \lya\ emission line with a FWHM $\simeq 500\mkms$ and a
  luminosity of $L_{\rm Ly\alpha} = 3.9 \sci{43} {\rm erg \, s^{-1}}$
  superposed on the trough of the PDLA.  Mechanisms for powering this
  large \lya\ luminosity are discussed. We argue against emission from
  \ion{H}{2} regions in the PDLA galaxy since this requires an
  excessive star-formation rate $\sim 500\,\msol \rm yr^{-1}$ and
  would correspond to the largest \lya\ luminosity ever measured from
  a damped \lya\ system or starburst galaxy.  We use a Monte Carlo
  radiative transfer simulation to investigate the possibility that the
  line emission is fluorescent recombination radiation from the PDLA
  galaxy powered by the ionizing flux of the quasar, but find that the
  predicted \lya\ flux is several orders of magnitude lower than
  observed. We conclude that the \lya\ emission is not associated with
  the PDLA galaxy at all, but instead is intrinsic to the quasar's
  host and similar to the extended \lya\ ``fuzz'' which is detected
  around many AGN. PDLAs are natural coronagraphs that block their
  background quasar at \lya\ , and we discuss how systems similar to
  SDSS~J124020.91$+$145535.6 might be used to image the neutral
  hydrogen in the PDLA galaxy \emph{in silhouette} against the
  screen of extended \lya\ emission from the background quasar.
\end{abstract}

\keywords{Galaxies: Evolution, Galaxies: Intergalactic Medium, 
Galaxies: Quasars: Absorption Lines}

\section{Introduction}

The ionizing radiation emitted by a luminous quasar can, like a
flashlight, illuminate hydrogen clouds in its vicinity, teaching us
about the size, kinematic structure, and density of the gas which
surrounds it \citep{rees88,hr01}. 
This is because recombinations from photoionized
hydrogen ultimately produce Ly$\alpha$ photons -- a fraction of the
energy in the quasar's UV continuum is `focused' into fluorescent line
radiation and re-emitted, allowing us to study the
physical conditions in the emitting gas.

Extended \lya\ nebulae with luminosities up to $L_{\rm
  Ly\alpha}=5\sci{42}~{\rm erg~s^{-1}}$ and angular sizes as large
$\sim 10\arcsec$ have been observed around many quasars
\citep[e.g.][]{djorgovski85,heckman91a,cjw+06}, 
but the physical mechanism powering
this emission is still unclear.  In addition to fluorescent emission
powered by photoionization, other mechanisms include emission from
material shocked by radio-jets or starburst outflows or \lya\ cooling
radiation from gravitational collapse.

The same set of physical mechanisms have been proposed to explain the
analogous \lya\ nebulae observed around a subset of luminous radio
galaxies \citep{msd+90,villar07}, as well
as the recently discovered \lya\ `blobs'
\citep{fmw99,steidel00,matsuda04}.  
The primary difference between the extended ``fuzz'' around
quasars as compared to the nebulae around radio galaxies and \lya\
blobs is that a luminous quasar and hence the source of ionizing
photons is directly detected in the former but not in the latter.
Obscuration and orientation effects, as are often invoked in unified
models of AGN, could be responsible for this difference. This
explanation is plausible considering that the large energies in the
jets $\gtrsim 10^{60}$\,erg \citep[e.g.][]{Miley80} of luminous radio
galaxies strongly suggest nuclear activity and evidence for an
obscured AGN has been uncovered in several of the \lya\ blobs
\citep{csw+04,dbs+05,gsc+07}.

In the course of a survey for damped \lya\ absorption proximate to
high $z$ quasars \citep[][see also Ellison et al. 2002 and Russell et
al. 2006]{phh08}, we discovered an extremely luminous \lya\ blob
coincident with both the redshift of a $z=3$ proximate damped \lya\
system (PDLA)\footnote{ A PDLA is defined as an absorber with $\mnhi
  \ge 2\sci{20} \cm{-2}$ located within $\delta v = 3000 \mkms$ of its
  background quasar} and its background quasar.  Although intervening
DLAs ($\delta v > 3000 \mkms$) rarely show \lya\ emission
\citep{scb+89,mff04,kwy+06}, PDLAs appear to preferentially exhibit
\lya\ emission superimposed on their \lya\ absorption trough
\citep{mw93,mwf98,eyh+02} which is $\sim 5$ times brighter than the
few intervening detections. Although based on poor statistics and
heterogeneous samples, this putative discrepancy suggests that the
\lya\ emission associated with PDLAs might be powered by its
background quasar. In what follows, we analyze the basic
characteristics of this absorber and discuss several physical
scenarios for the origin of this \lya\ blob.  Finally, we emphasize
several novel applications that exploit this unique configuration.
Throughout this paper we use the best fit WMAP3 cosmological model of
\citet{wmap03}, with $\Omega_m = 0.240$, $\Omega_\Lambda =0.76$,
$h=0.73$.  


\section{Observations}
\label{sec:lya}

\begin{figure}[ht]
  \centerline{\epsfig{file=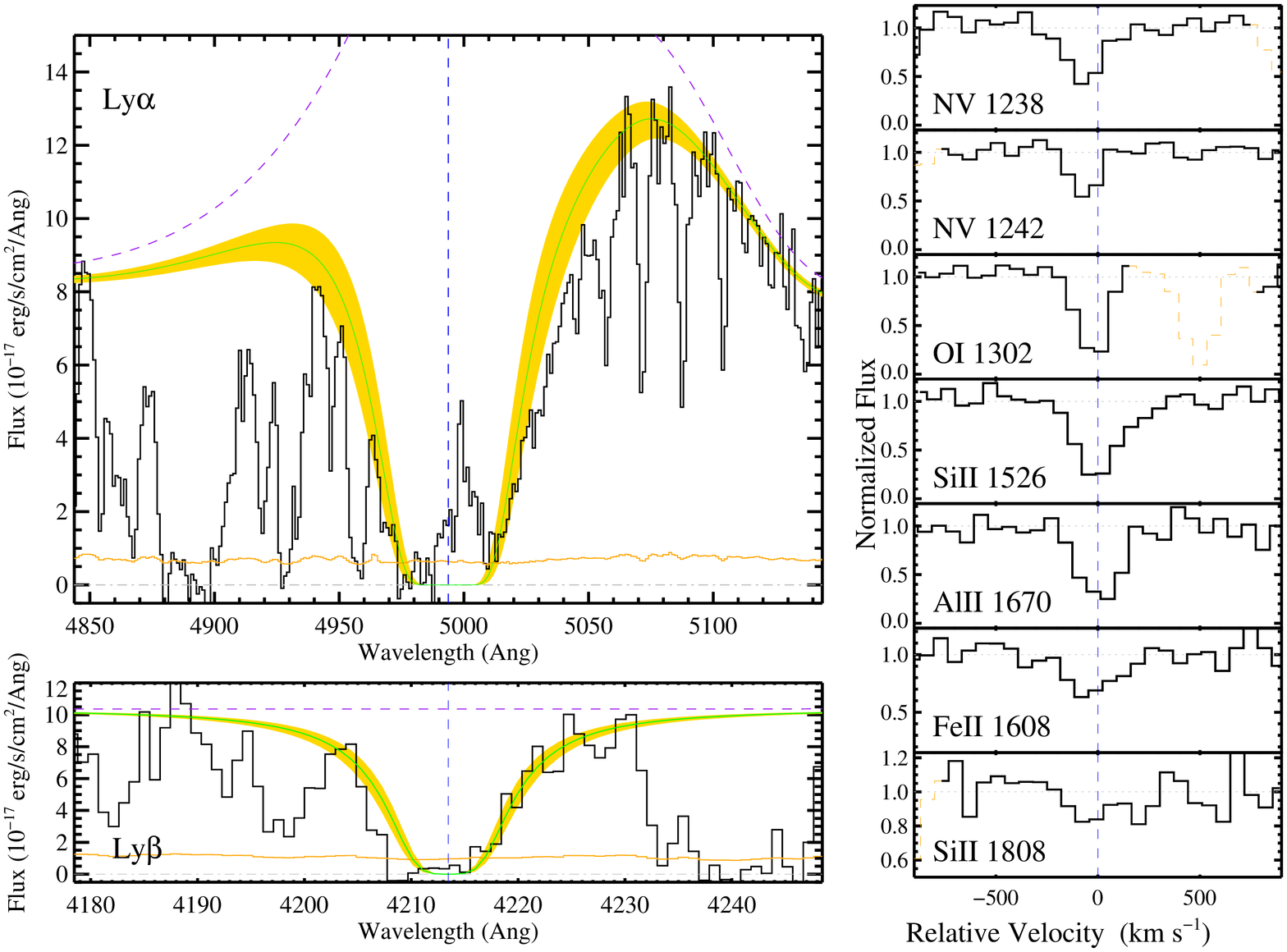,angle=90,width=0.45\textwidth}}
  \caption{ \lya, \lyb, and metal-line profiles for the proximate DLA
    system ($|\delta v| < 3000$km/s) at $z=3.1078$ toward \qso. 
    The absorption redshift is established by the
    low-ion transitions \ion{O}{1}~1302, \ion{Si}{2}~1526, and
    \ion{Al}{2}~1670 and confirmed by the \lyb\ profile.  Its \nhi\
    column density is measured from a Voigt
    profile fit to the damping wings of the \lya\ and \lyb\
    transitions, $\log \mnhi = 21.2 \pm 0.15$.   
    Near the center of core of the \lya\ profile one
    notices significant emission which we identify as \lya.
}
  \label{fig:metals}
\end{figure}


We recently completed a search of the Sloan Digital Sky Survey (SDSS),
Data Release 5 \citep[DR5;][]{sdssdr5} for damped \lya\ absorption at
small velocity separations from $z>2$ quasars \citep[PDLAs;][]{phh08}.
In the course of this PDLA survey, we identified a damped \lya\ system
with absorption redshift $z_{\rm abs} \approx z_{\rm qso}$ in the
spectrum of the radio-quiet quasar SDSS~J124020.91$+$145535.6
(henceforth \qso). The redshift of the quasar is $z_{\rm
  qso} = 3.1092 \pm 0.0014$, which was precisely determined from a
near-infrared spectrum of the H$\beta$ and [\ion{O}{3}] emission line
region (Hennawi et al., in prep), obtained with the Gemini Near
Infrared Imager and Spectrometer \citep[NIRI;][]{NIRI}.  In contrast to
the discovery of every other PDLA system, this absorber was initially
identified by the presence of strong metal-line absorption features
(Figure~\ref{fig:metals}) at a redshift $z_{\rm mtl} = 3.1078 \pm
0.0005$.  The large equivalent widths of these lines suggest an
optically thick \ion{H}{1} cloud at the Lyman limit and
correspondingly strong \lya\ absorption.

Contrary to our expectation, we observe significant flux in the
spectral region corresponding to the absorber's \lya\ transition.
Figure~\ref{fig:metals} shows the \lya\ and \lyb\ profiles of the
absorber and an estimate of the quasar flux.  Overplotted on the
data are Voigt profiles assuming a single \ion{H}{1} `cloud' with
$\mnhi = 10^{21.2 \pm 0.15} \cm{-2}$ and $z_{\rm abs} = z_{\rm mtl}$.
This model is a good representation of the data except for the flux in
the core of the \lya\ profile.  Because the metal-line transitions
and the \lyb\ data imply a damped \lya\ system, we are highly
confident that the emission at $\lambda \approx 5000$\AA\ is not
quasar continuum flux, but instead corresponds to line emission from
another source.  

The SDSS fiber spectra are spatially incoherent and only indicate that
the source of emission is within the 3$''$-diameter fiber.  We rule
out the identification of this line as [\ion{O}{2}] or any other
nebular line from a low redshift galaxy within the SDSS fiber because
there are no additional narrow emission lines in the SDSS spectrum and
furthermore the broad emission line profile does not resemble the [\ion{O}{2}]
doublet.  We fit the \lya\ emission with a Gaussian profile and find
FWHM~$=500 \pm 150$\kms\ with a centroid of $z_{Ly\alpha} = 3.113 \pm
0.001$, which is an offset of $\approx 400$\kms\ from $z_{\rm mtl}$
and $100 \pm 100 \mkms$ offset from $z_{\rm qso}$.  A simple boxcar
extraction of the line emission gives a flux $f_{\rm Ly\alpha} = 4.3
\pm 0.3 \sci{-16} {\rm erg \, s^{-1} \, cm^{-2}}$ and we derive a
luminosity $L_{\rm Ly\alpha} = 3.9 \pm 0.4 \sci{43} {\rm erg \, s^{-1}}$.

\cite{matsuda04} have provided a quantitative definition for \lya\
blobs as extended \lya\ emitting sources with isophotal areas larger
than $16~{\rm arcsec}^2$ and fluxes brighter than $f_{\rm Ly\alpha} >
0.7\sci{-16} {\rm erg~s^{-1}~cm^{-2}}$.  Their narrow band imaging
survey was also equivalent width limited to ${\rm EW_{obs}} > 80$~\AA.
The emission detected in the SDSS spectrum satisfies two of these
conditions.  Its flux would actually be the fifth brightest emitter in
the rich SSA22 protocluster field surveyed by \citet{matsuda04} and
its equivalent width exceeds ${\rm EW_{obs}} > 82$\AA\ based on the
non-detection of continuum flux in the core of the \lya\ profile
(i.e.\ at $\lambda \approx 4980$\AA).  To evaluate the spatial extent,
we acquired a short exposure of \qso\ with the Keck/LRIS
spectrometer and confirm it exceeds 16~arcsec$^2$.  We conclude that
this emission corresponds to a \lya\ blob.

A striking characteristic of the PDLA system is that it
exhibits a large \ion{N}{5} rest equivalent width, $W_\lambda({\rm NV
  1242}) = 0.30 \pm 0.05$\AA\ (Figure~\ref{fig:metals}).  
There are only a few examples in the literature of significant \ion{N}{5}
detections in a DLA \citep{lu96,fpl+07} and nearly all of these 
are proximate DLA with $z_{\rm abs} > z_{\rm qso}$ and
with associated \lya\ emission \citep{mw93}.
The large equivalent widths of the
\ion{N}{5} doublet for \name\
indicate the lines are saturated and we derive a
conservative lower limit to the column density of $\log \N{N^{+4}} >
14.4$\,dex.   We also infer a lower limit to the gas 
metallicity, based on the equivalent widths of low-ion absorption,
of 1/10 solar abundance.

\section{Discussion}
\label{sec:discuss}

We have discovered a \lya\ blob associated with a PDLA system at $z \sim 3$.
We will now discuss several scenarios for the origin of the \lya\ emission
which also imply distinct configurations for the PDLA system, quasar,
and \lya\ blob.

{\bf Star Formation from the PDLA galaxy:} Consider first that the
\lya\ emission arises from \ion{H}{2} regions associated with star
formation in the PDLA galaxy.  It is generally accepted that large
luminosities, large physical extent, and large equivalent width of the
line emission in \lya\ blobs cannot be the result of recombination
radiation in star forming regions in high-redshift galaxies.
Nevertheless, we estimate the star-formation rate required to power
the large \lya\ luminosity in \name.  We may place a lower limit on
the star-formation rate of $37\,\msol {\rm yr^{-1}}$ using the
\citet{kennicutt98} relation ${\rm SFR}(\msol~{\rm yr^{-1}}) = L({\rm
  H}\alpha)\slash 1.26\times 10^{41}~{\rm ergs~s^{-1}}$ and the case B
recombination value of the ratio $I({\rm Ly}\alpha)\slash I({\rm
  H}\alpha) = 8.3$.  This provides only a conservative lower limit
because corrections for dust extinction and/or absorption by
foreground \ion{H}{1} gas would imply a much larger SFR.  Indeed, the
\lya/H$\alpha$ ratio in LBGs \citep[measured from a small subset;][]{ess+06a} 
show an average of only $I({\rm
  Ly}\alpha)\slash I({\rm H}\alpha) =0.6$ with a standard deviation of
0.6.  This LBG \lya/H$\alpha$ ratio implies an even more extreme SFR~$
\sim 500\,\msol {\rm yr^{-1}}$.  Even our conservative lower limit
exceeds the SFR by an order of magnitude for all but one DLA
\citep{mff04,wgp05}.  And, our fiducial value assuming the LBG
\lya/H$\alpha$ ratio is nearly a factor of two larger than the largest
SFRs measured by \citet{ess+06a}. In addition, the \lya\ flux of
\name\ is larger than that of any of the 780 LBGs at $z\sim 3$ studied
by \citet{shapley03}.  Taking these arguments together, we conclude
that the \lya\ blob is not associated with star formation in the PDLA
galaxy.

\begin{figure}[ht]
  \centerline{\epsfig{file=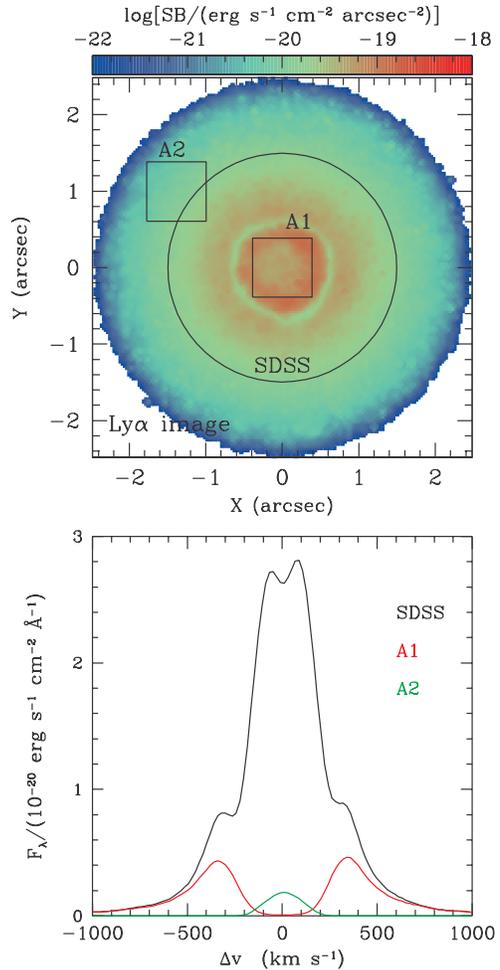,bb=0 139 300 720,
      width=0.4\textwidth}}
  \caption{Theoretical prediction for a PDLA at $z=3$.  The DLA is
    modeled as a singular isothermal sphere with dark matter halo mass
    of $10^{11} M_\odot$ and a quasar located directly behind it.  The
    top panel shows the predicted \lya\ image for this configuration
    taking into account self-shielding effects and the radiative
    transfer of fluorescent \lya\ photons.  The 1D spectra for two
    $0.73''\times 0.73''$ apertures (labeled A1 and A2) are shown in
    the lower panel where \lya\ at $z=3$ corresponds to 0km/s.  The 1D
    spectrum of the fluorescent radiation from the optically thick
    region of the cloud (A1) shows the characteristic double-peaked
    profile expected for \lya\ photons diffusing in frequency out of
    an optically thick medium.  The emission from the optically thin
    outer region of the cloud (A2) is a single peak centered at
    $\Delta v = 0 \mkms$ with much lower surface brightness.  Although
    the shape of the line profile for the SDSS aperture is in
    reasonable agreement with the observations, the predicted flux is
    several orders of magnitude fainter than observed.  }
  \label{fig:theory}
\end{figure}

{\bf Fluorescent Recombination Radiation Powered by the Quasar:} An
optically thick gas cloud illuminated by a nearby quasar will emit
Ly$\alpha$ photons at a rate which is $\eta \sim 0.6$ times the
ionizing flux impinging upon it \citep{gw96}.  If this is the
mechanism responsible for the \lya\ emission in \name, then one can
estimate its distance from the quasar as follows.  Suppose that the
emitting region has an angular diameter $\theta$ which covers a
fraction of the area of the $3\arcsec$ SDSS fiber, so that $f_{\rm
  Ly\alpha}= \Phi_{\rm Ly\alpha}\pi \theta^2\slash 4$, where
$\Phi_{\rm Ly\alpha}$ is the Ly$\alpha$ surface brightness.  At a
distance $d$ from the quasar, the DLA-galaxy absorbs an ionizing
photon flux of $F_{\rm c} = (1/4\pi d^2) \int (L_{\nu}/{h\nu})d\nu =
({1}/{4\pi d^2}) (L_{\nu_{\rm LL}}/{h\alpha_Q})$ where the last
equality assumes the quasar spectral energy distribution obeys the
power-law form $L_{\nu} = L_{\nu_{\rm LL}}(\nu\slash \nu_{\rm
  LL})^{-\alpha_{\rm Q}}$ blueward of the Lyman limit $\nu_{\rm
  LL}=13.6~{\rm eV}\slash h$, where $L_{\nu_{\rm LL}}$ is the specific
luminosity at the Lyman limit and $\alpha_{\rm Q}= 1.57$
\citep{telfer02}. We estimate $L_{\nu_{\rm LL}}=1.0\sci{31}{\rm
  erg}\,{\rm s}^{-1}\,{\rm Hz}^{-1}$, by combining the SDSS optical
photometry of the quasar \qso\ with the assumed
power-law spectral shape. The fluorescent surface brightness is given
by $\Phi_{\rm Ly\alpha} = \eta h\nu_{\rm
  Ly\alpha}\frac{F_c}{\pi}\frac{1}{ (1+z)^4}$, so we arrive at an
estimate for the distance $d =
186({\eta}/{0.60})({\theta}/{3\arcsec})\,{\rm kpc}$ adopting the
measured \lya\ flux.  The maximum possible distance is $d < 186\,{\rm
  kpc}$ because we have assumed (i) the fluorescent emission fills the
entire $3\arcsec$ SDSS fiber, and (ii) we have assumed a `full moon'
configuration (front-side illuminated) where our scenario is more
likely to approach a `new moon' (back-side) illumination pattern.  We
will address this latter issue in greater detail below.
  
Intriguingly, the above distance estimate of a few hundred kpc matches
the upper limit on the gas distance implied by the observed \ion{N}{5}
column density, if we assume the quasar is photoionizing the outer
layers of the PDLA system.  This is the most likely origin because the
extragalactic UV background and local sources are too weak or too soft
to yield such a large N$^{+4}$ column density 
and \ion{N}{5} is very rarely observed in DLA systems (Fox et al., in
prep).  We have estimated the distance of the quasar from the PDLA
system by performing a series of calculations with the Cloudy software
package \citep[v.6.0.2;][]{cloudy98} assuming a plane-parallel slab of
gas with $\mnhi = 10^{21.3} \cm{-2}$, metallicity [N/H]~$= -1$, and a
range of ionization parameters $U = F_c/n_H c$.  To match the observed
\ion{N}{5} equivalent width, we require $\log U > -2$ and, therefore,
an upper limit to the separation between the quasar and DLA of
$d_{{\rm NV}} < 700~{\rm kpc} (n_H/1~{\rm cm^{-3}})^{-1/2}$.  We draw two
important conclusions: (1) the gas is located within approximately
1\,Mpc of the quasar; (2) the majority of PDLA systems lying within
1\,Mpc of the quasar should also show significant \ion{N}{5}
absorption.

We explore the fluorescence scenario further by performing the
following numerical simulation.  The PDLA system is modeled as a
singular isothermal sphere within a halo of $10^{11}\msol$, for which
5\% of the mass is assumed to be in gas. The system is placed between
the observer and the quasar with a distance 300\,kpc to the quasar
such that the PDLA's backside is illuminated by the quasar. We use the
code developed in \citet[][see also Cantalupo et al. 2005]{Juna08} to perform
the self-shielding calculation, solving for the ionization structure
of the gas, the recombination rate, and the resulting \lya\
luminosity. The radiative transfer of the fluorescent \lya\ photons is
solved with a Monte Carlo method \citep{zheng02,Juna08}.  In turn, we
produce a map of the expected \lya\ surface-brightness as shown in
Figure~\ref{fig:theory}.  We have extracted several one-dimensional
spectra from this surface-brightness image to illustrate the flux
level and the expected emission line profiles.  Although the
one-dimensional spectrum through the $3''$ SDSS fiber aperture has
roughly the observed line width, it has an integrated flux that is
several orders of magnitude lower than observed.  The low flux results
from the fact that we are observing the illuminated galaxy in a `new
moon' configuration; the optically thick regions which give rise to
the \lya\ emission also serve to shield the radiation from our vantage
point.  The observed flux is instead dominated by the outer regions of
fiber which cover optically thin regions of gas in the halo
surrounding the galaxy (aperture A2 in Figure~\ref{fig:theory}).
Based on this calculation, we conclude that the observed \lya\
emission in \name\ is unlikely to be fluorescent recombination
radiation from the PDLA galaxy.

{\bf Extended \lya\ ``Fuzz'' from the Quasar Host Halo:} We consider
the most plausible hypothesis to be that the \lya\ emission is not
associated with the PDLA galaxy at all, but instead is intrinsic to
the quasar's host galaxy. Indeed, \lya\ emission has been detected as
an extended `fuzz' around many quasars
\citep[e.g.][]{djorgovski85,heckman91a,cjw+06}, 
and with luminosities comparable to
or greater than the $L_{\rm Ly\alpha} = 3.9\sci{43} {\rm erg \,
  s^{-1}}$ measured for \name. The velocity widths of this extended
emission can be as large as $1000-1500$~\kms consistent with our
observed kinematic profile. The physical mechanism responsible for
this emission could be fluorescent recombination radiation from gas in
the quasar host halo illuminated by quasar's large ionizing flux
\citep{rees88,heckman91a,hr01,weidinger04}.  It is important to
explain the distinction between the fluorescent \lya\ fuzz as it
occurs in the quasar host halo to the fluorescence from the DLA galaxy
which was considered in the previous section. For the former case the
emission is presumed to be coming from small dense self-shielding
clouds which permeate the quasar halo but have a small $\sim 0.5\%$
covering factor \citep{msd+90,heckman91a} deduced from the
fraction of the ionizing continuum which is being absorbed
\citep{heckman91a}. In the latter case, where a galaxy is illuminated,
the emission comes from a large, kpc-scale self-shielding cloud and a
large fraction $\eta \simeq 60\%$ of the impingent ionizing continuum
is converted into \lya\ recombination photons.  But the mechanism
powering the extended emission line `fuzz' could be among other
mechanisms put forth to explain the extended \lya\ halos around quasars
\lya\ blobs, and high-redshift radio galaxies, such as gas shock-heated
by a large scale outflow or cooling radiation from gravitational
infall \citep[e.g.][ and references therein]{matsuda04,ctf+06}.

In the halo fuzz scenario, the DLA-galaxy is in the foreground of the
extended \lya\ emitting region of the quasar host, and should hence
absorb the fraction of the \lya\ emission which it covers.  This
configuration also offers a reasonable explanation for why PDLAs might
preferentially exhibit \lya\ emission relative to intervening DLAs:
the presence of a PDLA occults the bright quasar making it much easier
to detect faint extended emission from the quasar halo.


\section{Future Directions and Applications}


We have reported on the discovery of a $z=3$ \lya\ blob associated
with a quasar that also exhibits a proximate damped \lya\ (PDLA)
system.  We argue that the line emission is not associated with star
formation in the PDLA galaxy nor is it fluorescent recombination
radiation from the backside of the PDLA gas illuminated by the
quasar. Instead, we conclude that this \lya\ emission is associated
with the halo of the quasar, and is similar to the extended \lya\
nebulae which have been observed around radio loud and radio quiet
quasars, high-z radio galaxies, and the \lya\ blobs. The thing that
makes the extended emission from \qso\ unique is that it
coincides with a proximate damped \lya\ system. 
Several other quasars 
show a similar coincidence of quasar, DLA, and extended
\lya\ nebula \citep[e.g.][]{mw93,mwf98,eyh+02}, yet \qso\ has
the largest \lya\ luminosity of them all.

With additional deep optical and near-IR spectroscopy one could detect
additional emission lines providing important constraints on the
mechanism powering the emission. Specifically, H$\beta$ lies in a
relatively clear transmission window near 2~microns as does H$\gamma$
near 1.8~microns. The relative flux of these Balmer lines to \lya\ is
sensitive to the amount of dust extinction, the hardness of the
photoionizing spectrum, and resonant scattering of the \lya\ photons.
These Balmer line ratios could
thus further distinguish between the emission mechanisms discussed in
\S~\ref{sec:discuss}. Furthermore, the detection of high excitation
lines, most notably \ion{N}{5}, \ion{C}{4}, or [\ion{He}{2}]~1640
would demand that the nebula is powered by a hard spectrum, because a
stellar continuum is too soft to produce strong emission in these
lines.

The coincidence of a damped \lya\ system and a quasar (PDLA) provide
unique opportunities for future research. A PDLA acts like a natural
coronagraph in the core of the damped \lya\ absorber, blocking the
background quasar's bright \lya\ emission and allowing one to conduct
a sensitive search for diffuse extended fuzz from the quasar, albeit
in a relatively narrow spectral window ($\Delta \lambda \approx
10-30$\AA).  At the faintest flux levels, star-formation from the PDLA
galaxy and/or the quasar host galaxy could contribute to the detected
emission, but for very bright \lya\ emission as in \qso\ it is very
unlikely that star-formation contributes significantly.

Because the bright \lya\ nebula likely originates in the quasar host
halo, and hence behind the PDLA, it is possible to image the PDLA
\emph{in silhouette} against the extended screen of \lya\ emission to
infer the spatial distribution of the PDLA \ion{H}{1} gas.  A `hole'
in the \lya\ emission will indicate the region where the \ion{H}{1}
gas is optically thick to \lya\ radiation, corresponding to to column
densities $\mnhi \gtrsim 10^{14} \cm{-2}$. The size of this hole would
thus indicate the extent of the lower column density `halo' of the
PDLA, and a seeing-limited, integral field unit (or narrow band)
observation would be sensitive to sizes of 5\,kpc and greater.  At
present, the only other means of assessing the size of absorption line
systems is by observing coincident absorption in close quasar pairs
\citep[e.g.][]{srs+95,ehm+07}, but this technique provides measurement
which are statistical in nature. Of the five other known PDLAs which
also exhibit extended \lya\ emission, two, PKS~0528$-$250 \citep{mw93}
and PHL~122 \citep{fmw99}, have been imaged with narrow band filters
centered on the cores of the DLAs. These observations were probably
too shallow to rule out the detection of the PDLAs \lya\
shadow. Furthermore, for PHL~1222 the significant contamination from
quasar \lya\ emission would have made the detection of a shadow even
more difficult. Future IFU or narrow-band imaging observations with
8-10m class telescopes could detect the \lya\ shadows of PDLAs
providing important constraints on the size and nature of these
absorbers.



\acknowledgments We acknowledge helpful discussions with B.
Matthews, P. Madau, D. Osterbrock, A. Gould, and J.
Miralda-Escud\'e. We are grateful to A. Shapely and D. Erb for
providing the distribution of LBG \lya\ and H$\alpha$ fluxes and for
helpful discussions.  JFH was supported by a NASA
Hubble Fellowship grant \# 01172.01-A 
and an NSF Postdoctoral Fellowship program (AST-0702879).  
JXP acknowledges funding through an
NSF CAREER grant (AST-0548180) and NSF grant (AST-0709235).  JK is
supported by NASA Hubble grant's HF-01197. ZZ gratefully acknowledges
support from the Institute for Advanced Study through a John Bahcall
Fellowship.



\end{document}